\documentstyle[12pt]{article}
\begin{document}
\title{ON CERTAIN RELATIONSHIPS BETWEEN COSMOLOGICAL OBSERVABLES IN
THE EINSTEIN-CARTAN GRAVITY}        
\author{D. PALLE \\Department of Theoretical Physics, Rugjer Bo\v skovi\'c 
Institute, \\P.O.Box 1016, Zagreb, CROATIA}
\date{ }
\maketitle
{\bf Summary.} - We show that  
 if one takes into account Penrose's conformal 
technique
in the Einstein-Cartan gravity, 
it is possible to obtain a relation between Hubble's expansion and Birch's
rotation of the Universe at spacelike infinity.
This relation also leads to the final
reduction of the gravitational coupling to a dimensionless quantity of
order unity at spacelike infinity. Initial-value equations at spacelike 
infinity in the Einstein-Cartan gravity can be used as a starting point to
study the structure formation and evolution of the Universe with $\Omega$=1.
Current measurements of the expansion and rotation of the Universe favour
the massive spinning particle as a candidate particle for the dark matter
.
  \\

Only a few years after Einstein (1915) had formulated his general theory of
relativity , Cartan (1922) realized the possible importance of torsion
for the geometrical theory of gravity.
However, only forty years later Kibble\cite{r1} and
Sciama\cite{r2} proposed a dynamical definition of spin connected with torsion,
and this theory was further elaborated by Hehl\cite{r3} and Trautman\cite{r4}.
Thus, in the Einstein-Cartan(EC) theory, the energy momentum and spin of matter are
coupled to the curvature and torsion of spacetime, respectively\cite{r1,r2}:

\begin{equation}
\begin{array}{c}
Q^{\rho}_{\mu\nu}\ =\ \frac{1}{2}(\Gamma^{\rho}_{\mu\nu}-\Gamma^{\rho}_{\nu\mu})\ =\ 
torsion \ tensor,\nonumber \\
c^{\rho}_{\mu\nu}\ =\ Q^{\rho}_{\mu\nu}+\delta^{\rho}_{\mu}Q_{\nu}
-\delta^{\rho}_{\nu}Q_{\mu}\ =
\ contortion\ tensor,\nonumber\\
Q_{\rho}\ =\ Q^{\mu}_{\rho\mu},\ \kappa\ =\ 8{\pi}G_{N}/c^{4},\ 
S^{\rho\mu\nu}\ =\ spin\ tensor,\nonumber \\
c^{\rho\mu\nu}\ =\ \kappa{S^{\rho\mu\nu}},\nonumber\\
\stackrel{\approx}{R}=Riemann-Cartan\ curvature\ scalar,\ h^{\mu}_{a}=tetrad,
\nonumber\\
T^{\mu}_{a}=energy-momentum\ tensor,\ \Lambda=cosmological\ constant,\nonumber\\
\stackrel{\approx}{R_{a}^{\mu}}-\frac{1}{2}h^{\mu}_{a}\stackrel{\approx}{R}\ 
=\ \kappa(T^{\mu}_{a}+h^{\mu}_{a}\Lambda).
\nonumber
\end{array}
\end{equation}

During the development of the EC theory it was observed that
one could construct nonsingular\cite{r5} and causal\cite{r6,r7} cosmological solutions
, contrary
to the singular and noncausal solutions in Einstein's gravity (due to 
the Penrose-Hawking-Geroch singularity theorems\cite{r8} and G\"{o}del's
noncausal solution\cite{r9}).
Although all EC solutions\cite{r5} are not nonsingular, it might be interesting
to study some simple but realistic cosmological models with baryonic matter and
background radiation. Owing to the algebraically coupled spin and torsion in the EC
gravity, Hehl showed\cite{r3} that all spin-torsion effects could be accounted for
by a new effective energy-momentum tensor.
The Raychauduri type of equation\cite{r5} with expansion, shear and vorticity for 
the EC theory is the simplest way to study scales within the Weyssenhoff fluid
model:

\begin{equation}
\begin{array}{c}
\kappa(\rho_{\gamma}+\rho_{B})\ =\ \frac{3}{R^{2}}(\stackrel{.}{R}
^{2}+k)+\frac{\kappa^{2}}{4}<S^{2}>,\\
\rho^{0}_{\gamma}/\rho_{\gamma}=R^{4}/R^{4}_{0},\ \rho^{0}_{B}/
\rho_{B}=R^{3}/R^{3}_{0},\nonumber\\
\rho^{0}_{\gamma},\rho^{0}_{B},R_{0}=present\ values,\ k=\pm1,0,\nonumber\\
R(t)\ =\ cosmological\ scale\ factor,\nonumber\\
<S^{2}>\ =\ averaged\ squared\ spin\ density,\nonumber\\
baryon\ number\ density=n_{B}=\frac{\rho_{B}}{m_{B}}=\frac{<S^{2}>^{1/2}}{\hbar},
\nonumber\\
assumption:\ {\stackrel{.}{H}}_{0}=0;\ H_{0}=\frac{\stackrel{.}{R}}{R}\ .\nonumber
\end{array}
\end{equation}

In the Weyssenhoff fluid we neglect contributions to shear
or vorticity, except the contribution to the effective vorticity from the torsion 
which is due to the baryon spin density.
Applying the extremality condition to equation (1), one arrives at
the algebraic equation whose solution gives us the minimum 
of the cosmological scale function:

\begin{equation}
\begin{array}{c}
{\gamma}x^{4}+x^{3}+{\alpha}x^{2}-\beta=0,\ \ \gamma=0,\mid\gamma\mid,
-\mid\gamma\mid\ for\ k=0,-1,+1,\nonumber\\
R_{min}\simeq{R_{0}}x_{0},\ x_{0}=(\beta/\alpha)^{1/2},\ for\ k=0,\pm1,\nonumber\\
R_{min}=\frac{1}{3}\frac{\hbar\rho^{0}_{B}}{cH_{0}m_{B}}(\frac
{3{\pi}G_{N}}{2\rho^{0}_{\gamma}})^{1/2},\ \\ 
\alpha=\frac{4}{3}\frac{\rho^{0}_{\gamma}}{\rho^{0}_{B}},\ 
\beta=\frac{{\pi}G_{N}}{2}\frac{\hbar^{2}\rho^{0}_{B}}{m_{B}^{2}},\ 
\mid\gamma\mid=\frac{1}{4{\pi}G_{N}R^{2}_{0}\rho^{0}_{B}}\ ,\nonumber\\
\rho^{0}_{B}{\simeq}(10^{-33}-10^{-32})g\ cm^{-3},\ \rho^{0}_{\gamma}{\simeq}
8{\times}10^{-34}g\ cm^{-3},\nonumber\\
H_{0}{\simeq}67\ km s^{-1} Mpc^{-1},\ R_{min}{\simeq}(10^{-16}-10^{-15})cm.
\nonumber
\end{array}
\end{equation}

For this simple cosmological model with 
a small fraction of the observed baryon density of stellar  
objects which pass
gravitational collapse and background radiation, the minimum of
the scale function is of the order of magnitude of the weak interaction
scale. This is the scale at which gauge, discrete and conformal symmetries in
particle physics are broken. The quantum principle enters here only at the
level of the first quantization, necessary to describe the spin densities
of matter fields. Torsion acts at small distances as a repulsive force
and dominates because of the stronger scaling $R^{-6}$. The energy conditions
of the singularity theorems\cite{r8} are violated by the modified        
energy-momentum tensor\cite{r5}.

This intriguing result can motivate us to see if we can improve the 
situation with G\"{o}del's noncausal solution. Actually, it was
observed a few years ago\cite{r6} that the EC theory had sufficient freedom
to exclude closed timelike curves. Obukhov and Korotky\cite{r7} 
later developed Weyssenhoff fluid theory in the EC gravity
and within their scheme successfully analysed the global rotation 
of the Universe observed by Birch\cite{r10}.

Their generalization of the G\"{o}del metric looks like this:

\begin{eqnarray}
ds^{2}\ =\ dt^{2}-R^{2}(dx^{2}+ka^{2}(x)dy^{2}+dz^{2})-2Rb(x)dydt,
\\
\ b(x)\ =\ \sqrt{\sigma}a(x),\ a(x)\ =\ Ae^{mx},\ R,k,\sigma,A,m=const.
\nonumber
\end{eqnarray}

The static solution of the EC field equations results in a completely causal
solution which might explain the observed global rotation of the
Universe\cite{r6,r7}. Our matter of concern is a note of Obukhov and Korotky
that the resulting torsion of spacetime equals the inverse of the radius of
the Universe. We want to show that this is not a coincidence.

The important concept that has been introduced is the hypersurface in the
neighbourhood of the infinity and the boundary hypersurface\cite{r11,r12,r13}.
On the other hand, the conformal transformation is the only one that
can connect and relate quantities at various scales. Thus, Penrose's
conformal technique\cite{r11} analyses the relation between the initial and 
transformed
values of tensors for conformally connected metrics\cite{r11,r14}:

\begin{equation}
\begin{array}{c}
{\tilde{P}}_{\mu\nu}\ =\ P_{\mu\nu}-\Omega^{-1}\nabla_{\mu}
\nabla_{\nu}\Omega+\frac{1}{2}\Omega^{-2}g_{\mu\nu}(\nabla_{\rho}
\Omega)(\nabla^{\rho}\Omega),\nonumber\\
P_{\mu\nu}\ =\ \frac{1}{2}R_{\mu\nu}-\frac{1}{12}R\ g_{\mu\nu},\ 
{\tilde{g}}_{\mu\nu}\ =\ \Omega^{-2}\ g_{\mu\nu}\nonumber
\end{array}
\end{equation}

and with a certain assumption on the energy-momentum tensor\cite{r11}:

\begin{equation}
\begin{array}{c}
\{\Omega^{2}({\tilde{R}}_{\mu\nu}-\frac{1}{2}\tilde
{R}{\tilde{g}}_{\mu\nu}-\kappa\Lambda{\tilde
{g}}_{\mu\nu})\}\ x^{\mu}x^{\nu}\ \leq\ 0\ ,\nonumber\\
\mbox{~~~}in\ the\ neighbourhood\ of\ infinity;
\mbox{~~~~}x^{\mu}=\ real\ vector.\nonumber
\end{array}
\end{equation}

For a positive definite $k/\sigma$, the static G\"{o}delian metric (3) fulfils 
the conditions of Penrose for asymptotically simple
spacetime\cite{r11}. The EC equations with this metric also describe 
the Cauchy initial-value problem. The solutions of
the equations with the matter-dominated Universe (vanishing pressure)
give the nonvanishing cosmological constant\cite{r7}, defined by the 
baryonic energy density and the observed vorticity of the Universe\cite{r10}. 

The consistency relation of Penrose gives us the relation between the
cosmological constant and the gradient of the cosmological scale
function at the boundary hypersurface\cite{r7,r11}:

\begin{equation}
\begin{array}{c}
-\kappa\Lambda\ =\ 3(\nabla_{\rho}\Omega)(\nabla^{\rho}\Omega),\nonumber\\
\Rightarrow\ -\kappa\Lambda\ =\ 3\ H^{2}_{\infty},\\
\sqrt{3}\ H_{\infty}\ \simeq\ {\mid}Q{\mid},\ for\ k/\sigma{\gg}1;\nonumber
\rho_{\infty}=-2\Lambda=\frac{8}{\kappa}\omega^{2}_{\infty}(\frac{k}{\sigma}
+1)^{2},\nonumber\\
Q=torsion=-\omega_{\infty}(2\frac{k}{\sigma}+1),\ \mid\omega_{\infty}\mid{\simeq}
\mid\omega_{0}\mid{\simeq}10^{-13}\ yr^{-1}.\nonumber
\end{array}
\end{equation}

It is interesting to note that the same relation can be derived in an empty
and flat de Sitter spacetime\cite{r15}, thus under physically unrealistic conditions.
The sign of the cosmological constant defines an asymptotic spacetime
as spacelike. The timelike gradient of the cosmological scale function,
orthogonal to the boundary hypersurface, is the Hubble constant by its
very definition. The relation (4) can be rewritten in the form
connecting Birch's rotation\cite{r10} and Hubble's expansion of the
Universe:

\begin{eqnarray}
H_{\infty}\ =\ \frac{2}{\sqrt{3}}(\frac{k}{\sigma}+1)\ \mid\omega_{\infty}{\mid}.
\end{eqnarray}

The modified EC energy-momentum tensor satisfies the necessary and sufficient
condition of non-negative definite modified energy density in the neighbourhood 
of the boundary hypersurface\cite{r11}. Thus the coincidence observed by
Obukhov and Korotky evidently has its deeper explanation acknowledging
conformal symmetry. It is important to point out that in the EC gravity the
existence of the global expansion requires global rotation and vice versa,
but eliminating the possibility of noncausal solutions $(k/{\sigma}=-1)$.

However, we have not exhausted all the consequences of the relation (4).
By direct insertions we immediately obtain

\begin{eqnarray}
G_{N}\ \rho_{\infty}\ H^{-2}_{\infty}\ =\ \frac{3}{4\pi}\ =\ {\cal O}(1).
\end{eqnarray}

This fact was known a long time ago\cite{r16}, but it was derived under 
ad hoc assumptions. We see that only the EC gravity can lead to
the final reduction of the gravitational coupling constant to a 
dimensionless number of order unity at spacelike infinity,
 because: (1) curvature to energy-momentum 
and spin to torsion are coupled with the same coupling constant,
(2) conformal mapping to spacelike infinity reduces the physical configuration
to a simpler one.

The effective critical density
of the Robertson-Walker expansion with the cosmological constant\cite{r15}
$({\rho}_{c})_{eff}=\rho_{c}+\Lambda$ is one half of the $\rho_{\infty}$
, thus $\rho_{c}=\rho_{\infty}$ and the Universe is open ($\Omega=1$).
The energy density of electromagnetic radiation scales as $R^{-4}$ (stronger
than the volume of the Universe expands), thus its contribution
vanishes at infinity.
At spacelike infinity pressures of baryonic and dark matter are expected to 
vanish exactly.
There should be a dominant contribution to torsion coming from
dark matter and it suggests that the particle candidate for dark
matter is a spinning particle (for example, a  massive neutrino).  
In that case, the relation (5) could be more easily satisfied by the presently
observed values of expansion and rotation\cite{r10}.

To conclude, we can say that the EC gravity significantly improves Einstein's 
general relativity explaining, without fine tuning, the connection between
cosmological and particle physics scales (Eq.(2)), the connection between
the expansion 
and rotation of the Universe (Eq.(5)) and the reduction of the gravitational
coupling to a number of order unity at spacelike infinity (Eq.(6)).
 Present cosmological measurements are only of the order of magnitude, but
new refined observations and establishing the handedness(chirality) of the rotation
of the Universe and its comparison with the handedness of the weak interaction
could be another challenge for astronomers. Finally,
in the light of the importance of conformal transformation, which is necessary 
to measure the whole Universe, 
the fact that the physical
 four-dimensional spacetime is the first low-dimensional spacetime with the
nonvanishing Weyl's conformal tensor\cite{r14}, is not incidental.

ACKNOWLEDGEMENT\\
I should like to thank the Commission of the European Union, Directorate
 General for Science, Research and Development (Brussels, Belgium) for partial
financial support under the contract CI1$^{*}$-CT91-0893 (HSMU).

\end{document}